\documentclass[twocolumn,amsmath,amssymb,pra]{revtex4-1}

\usepackage{graphicx}
\usepackage{color}
\usepackage{dcolumn}
\usepackage{amsmath}
\usepackage{CJK}
\usepackage{subfigure}
\usepackage{sidecap}

\usepackage{mathrsfs}
\usepackage{amsfonts}
\usepackage{indentfirst}
\usepackage{bm}

\newcommand{\be}{\begin{equation}}
\newcommand{\ee}{\end{equation}}
\newcommand{\bey}{\begin{eqnarray}}
\newcommand{\eey}{\end{eqnarray}}
\newcommand{\bw}{\begin{widetext}}
\newcommand{\ew}{\end{widetext}}

\newcommand{\ra}{\rangle}
\newcommand{\la}{\langle}

\newcommand{\ba}{\begin{array}}
\newcommand{\ea}{\end{array}}
\newcommand{\bi}{\begin{itemize}}
\newcommand{\ei}{\end{itemize}}
\newcommand{\bem}{\begin{enumerate}}
\newcommand{\eem}{\end{enumerate}}

\begin{document}

\title{Excited-state quantum phase transition and the quantum speed limit time}

\author{Qian Wang$^{1}$\footnote{Electronic address: qwang@zjnu.edu.cn}}

\affiliation{$^{1}$Department of Physics, Zhejiang Normal University, Jinhua 321004, China}

\author{Francisco P\'erez-Bernal$^{2,3}$\footnote{Electronic address: curropb@uhu.es}}

\affiliation{$^{2}$Depto.\ Ciencias Integradas y Centro de Estudios Avanzados en F\'{\i}sica, Matem\'aticas y Computaci\'on (CEAFMC). Universidad de Huelva, Huelva 21071, Spain\\
$^{3}$Instituto Carlos I de F\'{\i}sica Te\'orica y Computacional, Universidad de Granada, Granada 18071, Spain}

\date{\today}

\begin{abstract}
  We investigate the influence of an excited-state quantum phase
  transition on the quantum speed limit time of an open quantum
  system. The system consists of a central qubit coupled to a spin
  environment modeled by a Lipkin-Meshkov-Glick Hamiltonian. We show
  that when the coupling between qubit and environment is such that
  the latter is located at the critical energy of the excited-state
  quantum phase transition, the qubit evolution shows a remarkable
  slowdown, marked by an abrupt peak in the quantum speed limit time.
  Interestingly, this slowdown at the excited-state quantum phase
  transition critical point is induced by the singular behavior of the
  qubit decoherence rate, being independent of the Markovian nature of
  the environment.
\end{abstract}

\maketitle

\section{Introduction}

Quantum phase transitions are zero-temperature phase transitions in
which a system ground state undergoes a qualitative variation once a
Hamiltonian control parameter (such as an internal coupling constant
or an external field strength) reaches a critical value. Quantum phase
transitions are non-thermal and driven by quantum fluctuations in a
many-body quantum system \cite{Carr2010,Sachdev2011,Dutta2015}. They
are the best-known example of criticality in quantum mechanics and, in
the past decades, theoretical \cite{Emary2003, Zurek2005,
  Quan2006,Silva2008,Dziarmaga2010,Polkovnikov2011} and experimental
\cite{Greiner2002,Bloch2008,Baumann2010} groups have paid much
attention to this subject. Nowadays, quantum phase transitions are
considered a keynote topic in many fields of quantum physics. In
particular, ground state quantum phase transitions between different
geometrical limits of algebraic models applied to nuclear and
molecular structure have been extensively studied \cite{Gilmore1979,
  Feng1981, Iachello2004, Cejnar2007, Cejnar2010}.

More recently, it has been found that the quantum phase transition
concept, originally defined considering the system ground state, can
be generalized and extended to the realm of excited states
\cite{Cejnar2006,Cejnar2008,Caprio2008}. A characteristic feature of
excited-state quantum phase transitions is a discontinuity
in the density of excited states at a critical value of the energy of
the system. Such discontinuity, occurring at values of the control
parameter other than the critical one, is a continuation at higher
energies of the level clustering near the ground state energy that
characterizes the ground state quantum phase transition
\cite{Caprio2008,Leyvraz2005,PFernandez2009,Stransky2014}. Establishing
a parallelism with the Ehrenfest classification for ground state
quantum phase transitions, excited-state quantum phase transitions can
also be characterized by the nonanalytic evolution of the energy of an
individual excited state of the system when varying the control
parameter \cite{Caprio2008,Relano2008,PBernal2008,PFernandez2011}.

During the last decade, excited-state quantum phase transitions have been theoretically analyzed in
various quantum many-body systems: the Lipkin-Meshkov-Glick (LMG)
model \cite{Cejnar2007,PFernandez2009}, the vibron model
\cite{Caprio2008,PBernal2008}, the interacting boson model
\cite{Cejnar2009}, the kicked-top model \cite{Bastidas2014}, and the
Dicke model \cite{PFernandez2011,PFernandez2011b, Brandes2013,
  Kloc2017} among others.  Furthermore, the existence of a relation
between excited-state quantum phase transitions and the onset of chaos has been explored in
Ref.~\cite{PFernandez2011b} and excited-state quantum phase transition signatures have been
experimentally observed in superconducting microwave billiards
\cite{Dietz2013}, different molecular systems
\cite{Larese2011,Larese2013}, and spinor Bose-Einstein condensates
\cite{Zhao2014}.  The excited-state quantum phase transition influence on the dynamics of quantum
systems has recently attracted significant attention and several
remarkable dynamical effects of excited-state quantum phase transitions have been revealed
\cite{Relano2008,PFernandez2009,PFernandez2011,Santos2015,Santos2016,PBernal2017,Wang2017,
Engelhardt2015,Puebla2013,Puebla2015,Kloc2018}.
In particular, the impact of excited-state quantum phase transitions on the adiabatic dynamics of a
quantum system has been recently analyzed \cite{Kopylov2017}, as well
as the relationship between thermal phase transitions and excited-state quantum phase transitions
\cite{PFernandez2017}.

On the other hand, due to fast progresses in the fields of quantum
computation and quantum information science, the achievement of fast and
controlled evolution of quantum systems is a need of the hour. As a
fundamental bound,  imposed by
quantum mechanics, on the evolution speed of a system, the quantum speed limit time is in the spotlight
and its study has drawn considerable attention both in isolated and open
quantum systems
\cite{Deffner2013b,Taddei2013,delCampo2013,Zhang2014,Sun2015,Liu2016,Marvian2016,
  Ektesabi2017,Xiangji2017} (many more references can be found in the
recent reviews \cite{Frey2016,Deffner2017}).  The quantum speed limit
time, whose origin can be traced back to the reinterpretation of the
Heisenberg time-energy uncertainty relation by Mandelstam and Tamm
\cite{Tamm1945}, is defined as the minimum evolution time required by
a quantum system to evolve between two distinct states
\cite{Brody2003,Deffner2013b,Frey2016}.  The quantum speed limit time
sets a lower bound to the time needed to evolve between two
distinguishable states of a given system, and thus provides qualitative information
about the time evolution of a system without explicitly solving the system dynamics \cite{Peter1995}.

Nowadays, the quantum speed limit time is a quantity that has a
growing importance in quantum physics. In fact, it has been used to
set fundamental limits to the speed of computing devices
\cite{Lloyd2000}, to the performance of quantum control
\cite{Caneva2009}, and to the entropy production rate in
nonequilibrium quantum thermodynamics \cite{Deffner2010}; to best
parameter estimation in quantum metrology
\cite{Giovannetti2011,Braun2018}; and to analyze information
scrambling \cite{Campo2017}. See \cite{Frey2016, Deffner2017} and
references therein for a more exhaustive list of applications of the
quantum speed limit.  Moreover, the quantum speed limit may help us
figuring out how to control and manipulate quantum coherence
\cite{Xiangji2017}. 
For a given evolution time, the ratio between the quantum speed limit time
and the driving time provides an estimation of the potential capacity for speeding up the quantum dynamic evolution. Namely, a ratio equals to $1$ indicates that there is no room for further
 acceleration, but if the ratio is less than $1$, the smaller the ratio, the larger the possible system evolution
 speed up \cite{Deffner2017, CLiu2015}.
Interestingly, there is a common belief that the
quantum speed limit is a strictly quantum phenomenon, without a
classical counterpart. However, speed limit bounds in classical
systems, in close connection with the quantum speed limit, have been
recently reported \cite{Shanahan2018,Okuyama2018}.

For open quantum systems, one can expect that decoherence effects that stem from the
interaction between the system and the environment will strongly influence the quantum speed limit. It has already
been shown that the system decoherence can be enhanced by the occurrence of quantum
criticality in the environment \cite{Quan2006,Relano2008}.  Hence, one
would naturally expect an important variation in the quantum speed limit time whenever the environment undergoes a quantum phase transition.  Indeed, recent
results obtained for open systems consisting of a qubit coupled to a
spin \(\frac{1}{2}\) XY chain with nearest neighbors interaction \cite{Wei2016} or
to a LMG bath \cite{Hou2016} clearly verify that the quantum speed limit time has a
remarkable variation in the critical point of the environment ground
state quantum phase transition. This happens to such an extent that the quantum speed limit time has been
proposed as a possible probe to detect the occurrence of ground state
quantum phase transitions \cite{Wei2016,Hou2016}.

In this work, we extend the results obtained for ground state quantum
phase transitions in Refs.~\cite{Wei2016,Hou2016} to excited states
and excited-state quantum phase transitions; to study the
relationship between criticality in excited-state quantum phase
transitions and the quantum speed limit. With this aim, we analyze the
quantum speed limit time of an open system that consists of a central
qubit coupled to a LMG environment \cite{Lipkin1965}.  The net effect
of the coupling between the central spin and its environment is a
change of the environment Hamiltonian control parameter
\cite{Cucchietti2007,Relano2008,PFernandez2009,Rossini2007}.
Therefore, the variation of the coupling strength can drive the
environment through the critical point of an excited-state quantum
phase transition. In this case we report the effects that the crossing
has on the quantum speed limit time.  Furthermore, since the Markovian
nature of the environment has a strong influence on the properties of
the quantum speed limit, we will also study how the excited-state
quantum phase transition affects the environment's Markovian nature,
looking for a deeper understanding on the relationship between
excited-state quantum phase transitions and the quantum speed limit
time.

The rest of this article is organized as follows.  In Sec.~\ref{QSL},
we introduce the quantum speed limit for open systems and shortly
review some key concepts, due to the nontrivial way
of addressing the concept of quantum speed limit for such systems.  In
Sec.~\ref{SMS}, we describe the model, consisting of a central qubit
coupled to the LMG model, and analyze the phase transitions of the
environment, especially those related to excited-state quantum phase
transitions.  In Sec.~\ref{Results} we define the qubit quantum speed
limit time and investigate how the excited-state quantum phase
transition affects this quantity.  Sec.~\ref{PMc} is devoted to the
search of a physical explanation for the singular behavior of the
quantum speed limit time at the critical point of the excited-state
quantum phase transition.  We discuss in this section how the
variation of the quantum speed limit time at the excited-state quantum
phase transition critical point stems from the singular behavior of
the qubit decoherence rate and not from the Markovian nature of the
environment.  Finally, we discuss and summarize our results in
Sec.~\ref{Final}.  We have included an Appendix where we prove several
equations that are used in the main text.

 \begin{figure}
  \includegraphics[width=\columnwidth]{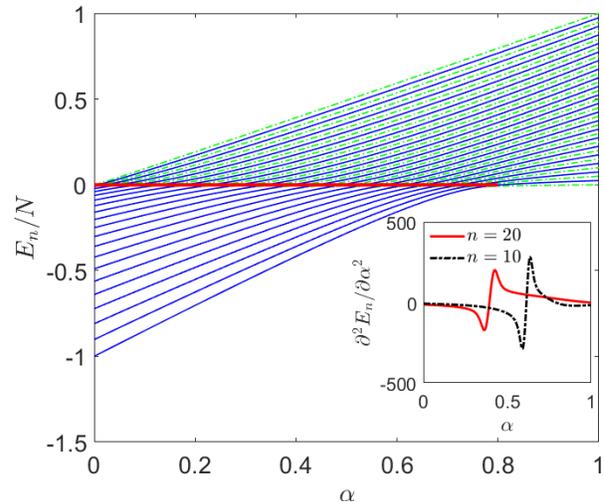}
  \caption{Energy levels of the environment Hamiltonian (\ref{LMGH}) as a function of the control parameter $\alpha$ with $N=40$.
   Odd (even) parity levels are depicted with blue solid (green dot-dashed) curves.
   The thick red line indicates the critical energy, $E_c=0$, of the excited-state quantum phase transition.
   Inset: The second derivative of the level energy with respect to the variable $\alpha$ for the $10$th and $20$th excited states, respectively.}
  \label{Energys}
 \end{figure}

\section{The quantum speed limit in open systems} \label{QSL}

In this section we briefly review the key concepts of the quantum
speed limit in open systems and establish the notions and the notation
that will be used in the rest of this article. We follow the approach
presented in Refs.~\cite{Deffner2013b,Deffner2017}, where the quantum speed limit
for a driven open system is obtained using a geometric
approach.

Consider a driven open quantum system with initial state
$\rho_0$, its evolution is governed by the quantum master equation
\cite{Deffner2013b,Breuer2007}

\be
\label{Dyopen}
\dot{\rho}_t=L_t(\rho_t),
\ee 

\noindent where $L_t$ is an arbitrary Liouvillian superoperator and
the dot denotes the time derivative.  The quantum speed limit time is
derived as a lower bound to the evolution time between an initial
state $\rho_0$ and a final state $\rho_t$, which can be obtained with
Eq.~(\ref{Dyopen}).  In the geometric approach the distance between
two quantum states is measured by the Bures angle
\cite{Deffner2013b,Jozsa1994}

\be
\label{BAD}
\mathcal{L}(\rho_0,\rho_t)=\arccos\left(\sqrt{F(\rho_0,\rho_t)}\right),
\ee

\noindent where
$F(\rho_0,\rho_t)=[\mathrm{tr}(\sqrt{\rho_t}\rho_0\sqrt{\rho_t})]^2$
is the quantum fidelity between $\rho_0$ and $\rho_t$.  According to
Eq.~(\ref{BAD}), the Bures angle is a measure of the distance between
states and the quantum speed limit can be interpreted as the the
maximum possible speed to sweep out the angle
$\mathcal{L}(\rho_0,\rho_t)$ under the dynamics governed by
Eq.~(\ref{Dyopen}) \cite{Shanahan2018}. The expression of the quantum
speed limit, $\nu$, can therefore be obtained by taking the time
derivative of the Bures angle \be \nu =
\dot{\mathcal{L}}(\rho_0,\rho_t)\leq|\dot{\mathcal{L}}(\rho_0,\rho_t)|.
\ee This is a cumbersome operation in the general case, as it involves
square roots of operators, but using the definition (\ref{BAD}) and
after some algebra, the above inequality can be written as
\cite{Deffner2013b}.
\be
\label{Ieq}
2\cos[\mathcal{L}(\rho_0,\rho_t)]\sin[\mathcal{L}(\rho_0,\rho_t)]
\dot{\mathcal{L}}(\rho_0,\rho_t)\leq|\dot{F}(\rho_0,\rho_t)|.
\ee

For
an initial state that is pure, $\rho_0=|\psi_0\ra\la\psi_0|$, and making
use of the von Neumann and Cauchy-Schwarz inequalities, inequality
(\ref{Ieq}) can be further simplified as \cite{Deffner2013b}
\begin{align} \label{QVI}
  2\cos[\mathcal{L}(\rho_0,\rho_t)]\sin[\mathcal{L}(\rho_0,\rho_t)]
  \dot{\mathcal{L}}(\rho_0,\rho_t)\leq|\la\psi_0|\dot{\rho}_t|\psi_0\ra| \notag \\
   \leq\mathrm{min}\{||L_t(\rho_t)||_1, ||L_t(\rho_t)||_2, ||L_t(\rho_t)||_\infty\},
\end{align}
where Eq.~(\ref{Dyopen}) has been used to get the second inequality and $||\mathcal{B}||_p=\left(
\sum_k b_k^p\right)^{1/p}$ with $b_k$ is the $k$th singular value of $\mathcal{B}$ denoting the Schatten $p$ norm of the operator $\mathcal{B}$. The cases $p=1, 2, \infty$  correspond with the trace norm, the Hilbert-Schmidt norm, and the operator norm, respectively.

Integrating Eq.~(\ref{QVI}) over time from $t=0$ to $t=\tau_e$, we can obtain the quantum speed limit time $\tau_e\geq\mathrm{max}
\{\tau_1,\tau_2,\tau_\infty\}$ where 
$\tau_p=\sin^2[\mathcal{L}(\rho_0,\rho_{\tau_e})]/\Gamma_{\tau_e}^p$ with
$\Gamma_{\tau_e}^p=(1/\tau_e)\int_0^{\tau_e}dt||L_t(\rho_t)||_p$ and $p=1, 2, \infty$.
Then, a unified expression of the quantum speed limit time for generic open system
dynamics can be defined as 
\be \label{QSLT}
 \tau_{\mathrm{QSL}}=
\mathrm{max}\left\{\frac{1}{\Gamma_{\tau_e}^1}, \frac{1}{\Gamma_{\tau_e}^2}, 
\frac{1}{\Gamma_{\tau_e}^\infty}\right\}\sin^2\left[\mathcal{L}(\rho_0,\rho_{\tau_e})\right].
\ee
Here, we stress that the derivation of Eq.~(\ref{QSLT}) assumes a pure initial state and it cannot be
applied to mixed initial states.
Properly formulating the quantum speed limit time for mixed initial states is a delicate issue which is still under study \cite{Zhang2014,Sun2015,Pires2016,Ektesabi2017}. 

The quantum speed limit time was experimentally investigated in a cavity QED system via  
the second order intensity correlation function and observing the speed up of the system evolution under environment changes \cite{Cimmarusti2015}.
Nevertheless, a direct experimental estimation of the quantum speed limit time itself is still an open problem.
It is noteworthy that a possible procedure to test the quantum speed limit time 
in quantum interferometry has been recently proposed \cite{Mondal2016}.

\section{Model}  \label{SMS}

We consider a system composed by a qubit coupled to a spin
environment. The Hamiltonian of the total system is
\cite{Quan2006,Cucchietti2007,PFernandez2009} 
\be
\hat H=\hat H_{\mathcal{E}}+\hat H_{\mathcal{SE}}, 
\ee 
where $\hat H_{\mathcal{E}}$ is the
environment Hamiltonian and the interaction between qubit and
environment is modeled by $\hat H_{\mathcal{SE}}$, which can be written as
\cite{Relano2008,PFernandez2009}
\be \label{SEH}
\hat H_{\mathcal{SE}}=|0\ra\la0|\otimes \hat H_0+ |1\ra\la1|\otimes \hat H_1,
\ee
where, $|0\ra$ and $|1\ra$ denote the two components of the qubit,
while $\hat H_0$ and $\hat H_1$ are interaction Hamiltonians that act upon the
environment Hilbert space.  Therefore, depending on the particular
qubit state, the environment evolution takes place under a different
effective Hamiltonian
$\hat H_\mathcal{E}^l=\hat H_\mathcal{E}+\hat H_l$ with $l=0,1$. 
As we will see below, the effect of $\hat{H}_l$ is to change 
the strength of the second term in the
environment Hamiltonian (\ref{LMGH}). 
Therefore, varying the interaction Hamiltonian $\hat H_l$, one can drive the
environment through its critical point
\cite{Relano2008,PFernandez2009}.

Specifically, the environment in our study is given by the generalized
LMG model.  This model, originally introduced as a toy model in
nuclear structure to test the validity of different approximations
\cite{Lipkin1965}, has been recently extensively used to study excited-state quantum phase transitions
\cite{Relano2008,PFernandez2009,Santos2016,Sindelka2017} and has
been implemented in the laboratory making use of different platforms: trapped
ions \cite{Monz2011, Lanyon2011}, large-spin molecules \cite{Gatteschi2003}, Bose-Einstein condensates \cite{Albiez2005, Gross2010,Riedel2010}, optical
cavities \cite{Leroux2010}, and cold atoms \cite{Makhalov2019}. The LMG model Hamiltonian is
\be
 \label{LMGH}
 \hat H_\mathcal{E}=-\frac{4(1-\alpha)}{N}\hat
 S_x^2+\alpha\left(\hat S_z+\frac{N}{2}\right),
 \ee
 where $N$ is the size of the environment, the control
 parameter $0\leq\alpha\leq1$ denotes the strength of the magnetic
 field along the $z$ direction, and
 $\hat{S}_\gamma=\sum_{j=1}^N\hat{\sigma}_\gamma^j$, the sum of Pauli spin
 matrices \(\hat{\sigma}_\gamma\) for $\gamma=\{x,y,z\}$.  

The total spin in the LMG model is a conserved quantity, 
i.e., $[\hat{H}_\mathcal{E},\hat{\mathbf{S}}^2]=0$ \cite{Dusuel2005}. In our study, 
only the maximum spin sector $S=N/2$ will be considered. Therefore, the dimension of the 
Hamiltonian matrix is  $\mathrm{Dim}[H]=N+1$.  
 Moreover, as the Hamiltonian (\ref{LMGH}) conserves parity, the parity operator,
 $\Pi=(-1)^{S+m}$ with $m$ is the eigenvalue of $\hat{S}_z$, split the Hamiltonian matrix into even- and odd-parity blocks, with dimension
 $\mathrm{Dim}[H_\mathcal{E}]_{\mathrm{even}}=N/2+1$ and
 $\mathrm{Dim}[H_\mathcal{E}]_{\mathrm{odd}}=N/2$ \cite{Santos2016}.
 In this work, we only consider even parity states, the subset that
 includes the ground state.  
 We should point out that for simplicity, we consider $\hbar=1$ throughout this article and set the quantities in our study as unitless.

 \begin{figure}
  \includegraphics[width=\columnwidth]{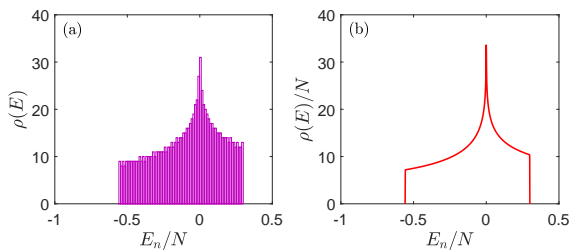}
  \caption{(a) Density of states of the environment Hamiltonian (\ref{LMGH}) for $N=2000$ with $\alpha=0.3$.
   (b) Normalized density of states of $\hat H_\mathcal{E}$, calculated by means of Eq.~(\ref{CDOS}), with $N=2000$ and $\alpha=0.3$.}
  \label{density}
 \end{figure}

 \subsection{Phase transitions in the environment}

 It is well known that the LMG model Hamiltonian in Eq.~(\ref{LMGH})
 exhibits a second-order ground state quantum phase transition
 at a critical value of the control parameter $\alpha_c=0.8$ \cite{Relano2008,PFernandez2009,Romera2014,Sindelka2017}.  The
 phase with $\alpha>\alpha_c$ is the symmetric phase, while the
 broken-symmetry phase corresponds to control parameter values
 $\alpha<\alpha_c$ \cite{PFernandez2009}.  However, besides the ground
 state quantum phase transition, the Hamiltonian (\ref{LMGH}) 
 also displays an excited-state quantum phase transition and
 the signatures of the ground state quantum phase transition, e.g., the nonanalytical
 evolution of the ground state as the control parameter of the system
 is varied,  propagates to  excited states for a control parameter value
 $\alpha<\alpha_c$, i.e.,  in the broken-symmetry phase.

 The LMG model correlation energy diagram is shown in
 Fig.~\ref{Energys}, depicting energy levels as a function of the
 control parameter $\alpha$ for $\hat H_\mathcal{E}$ with $N=40$.  In
 this figure it can be easily noticed how pairs of eigenstates with
 different parity are degenerate for $E<0$, and nondegenerate when
 $E>0$.  Moreover, the energy gap between adjacent energy levels
 approaches zero around $E\approx0$ in the broken-symmetry phase,
 where energy levels concentrate around $E=0$, marking the high
 local energy level density that characterizes excited-state quantum phase transitions. 
 Each excited energy level has an inflection point at $E\approx0$ \cite{Caprio2008} that induces 
 a singular behavior in the second derivative
 of every individual excited state energy with respect to the control
 parameter $\alpha$ (see the inset of Fig.~\ref{Energys})
 \cite{PBernal2008}.  A true nonanalytic behavior only happens in the large
 \(N\) limit, also called thermodynamic or mean field limit, but even
 for systems of finite size (\(N = 40\), as in Fig.~\ref{Energys}), there
 are clear precursors of the excited-state quantum phase transition effects, 
 like the behavior of $\partial^2E_n/\partial\alpha^2$ for $n>0$
 \cite{PBernal2008}. Something similar happens for the ground state
 energy, the $n=0$ case, that shows a discontinuity at the critical
 value of the control parameter, $\alpha_c$, associated with the ground state quantum phase transition. 
 The excited-state quantum phase transition can be crossed in two ways: 
 varying the energy for a fixed
 parameter value $\alpha < \alpha_c$ or choosing an
 excited state and varying the control parameter. It is important to
 understand that different excited states 
 cross the excited-state quantum phase transition at
 different values of the control parameter $\alpha$, as shown in the inset of Fig.~\ref{Energys}.

 For a given control parameter value $\alpha < \alpha_c$, the piling
 of energy levels at $E\approx0$ that marks the excited-state quantum phase transition will lead, 
 in the large \(N\) limit, to a discontinuity in the density of states.  
 Indeed, as shown in Fig.~\ref{density}(a), around $E=0$ the
 density of states $\rho(E)$ of $\hat H_\mathcal{E}$ has a peak for finite $N$,
 which will transform to a logarithmic divergence as $N\to\infty$
 \cite{Caprio2008, Ribeiro2008}. Therefore, the critical energy of the
 environment (\ref{LMGH}) is $E_c=0$ and, in fact, excited-state quantum phase transitions are
 often characterized by the singular behavior of the density of states occuring at
 the critical energy $E_c$, for fixed values of the control
 parameter  \cite{Caprio2008,PFernandez2009,Santos2016,Leyvraz2005,Stransky2014}.
The divergence of the density of states at $E_c=0$ can be understood using a semiclassical approach or the coherent state approach \cite{Caprio2008}.

 \begin{figure*}
  \includegraphics[width=0.85\textwidth]{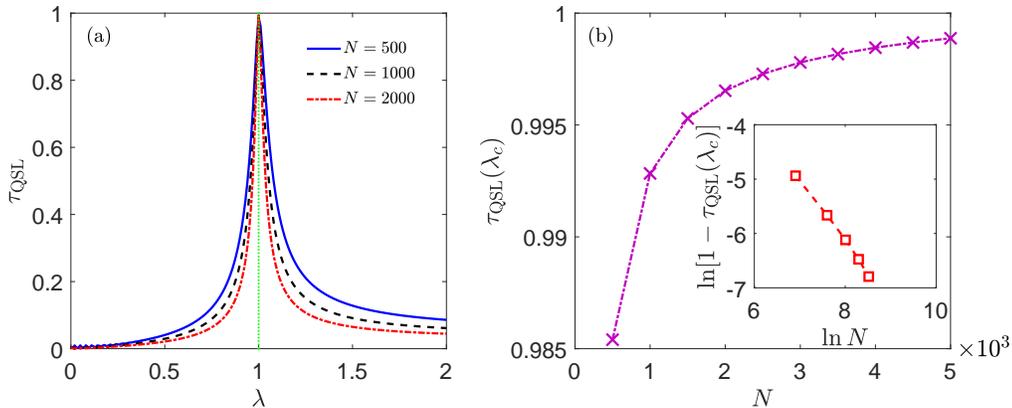}
  \caption{(a) The quantum speed limit time of the qubit as a function of the coupling strength
   $\lambda$ for different environment sizes (\(N\) values) for a system with a control parameter $\alpha=0.4$.
   The actual evolution time is set to unity, $\tau_e=1$.
   The vertical green dotted line indicates the critical value of the Hamiltonian parameter $\lambda_c$ obtained from
   Eq.~(\ref{LAMC}). (b) The critical quantum speed limit time $\tau_{\mathrm{QSL}}(\lambda_c)$ as a function of
   the environment size, $N$, with $\alpha=0.4$ and $\tau_e=1$.
   Inset: $1-\tau_{\mathrm{QSL}}(\lambda_c)$ as a function of $N$, in a double logarithmic scale.}
  \label{QSLa}
 \end{figure*}

 In the semiclassical limit, the quantum expression for the density of states 
 $\rho(E)=\sum_j(E-E_j)$ can be
 decomposed into a smooth and an oscillatory component \cite{Gutzwiller1990}, 
 $\rho(E)=\bar{\rho}(E)+\tilde{\rho}(E)$.
 The smooth part, $\bar{\rho}(E)$, is given by the classical phase space integral, 
 while the oscillatory term, $\tilde{\rho}(E)$,
 can be expressed as a sum over classical periodic orbits \cite{Gutzwiller1990}.
 In the classical limit, the oscillatory part can be omitted \cite{PFernandez2011}. 
 Then the density of states of the environment (\ref{LMGH}) has the following form
 \be \label{CDOS}
   \rho(E)=\bar{\rho}(E)=\left(\frac{N}{2\pi}\right)
                         \int\delta[E-\mathcal{H}(x,p)]dxdp,
 \ee
 where the environment's classical counterpart Hamiltonian
 $\mathcal{H}(x,p)$ can be obtained via the coherent or intrinsic
 state approach \cite{PFernandez2009,PFernandez2011}.  
Equation~(\ref{CDOS}) shows that the divergence
 in the density of states stems from the nonanalytical dependence of
 the classical phase space volume on the energy.  In addition,
 such nonanalyticity is usually associated with stationary points
 of the classical Hamiltonian  \cite{PFernandez2011,Stransky2014,Heiss2002}.

 In Fig.~\ref{density}(b), we plot the smooth part of the environment density of states
 for Hamiltonian (\ref{LMGH}) as a function of the eigenenergies.
 Note that the density of states has been normalized by the size of the environment.
 Obviously, the density of states exhibits a cusp singularity (i.e., an infinite peak) at $E_c=0$.
 Therefore, we confirm that the critical energy of 
 excited-state quantum phase transition is located at $E_c=0$.
 Moreover, a very good agreement between Figs.~\ref{density}(a) and \ref{density}(b) can be clearly appreciated.

 \begin{figure}
  \includegraphics[width=\columnwidth]{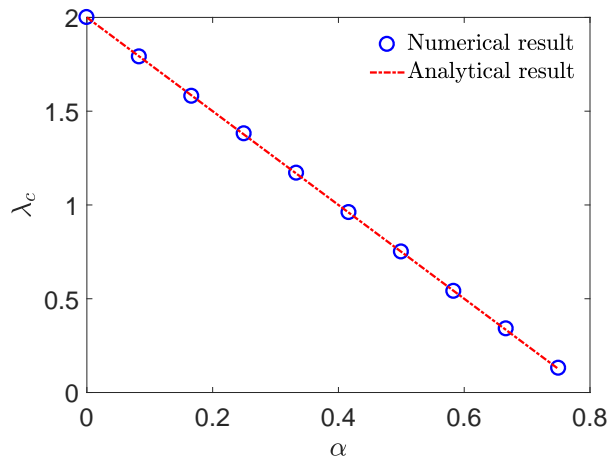}
  \caption{Critical coupling strength, $\lambda_c$, as a function of $\alpha$
   with $\tau_e=1$ and an environment size $N=1000$.
   The numerical result is obtained by identifying the location of the maximum point in $\tau_\mathrm{QSL}$,
   while the analytical result is provided by Eq.~(\ref{LAMC}).}
  \label{NALmc}
 \end{figure}

 \subsection{The critical coupling}

 In our study, the qubit is coupled to the environment (\ref{LMGH})
 with an interaction term
 \be
   \hat H_{\mathcal{SE}}=\lambda\sigma_\mathcal{S}^z \hat S_z,
 \ee
 \noindent where $\lambda$ is the coupling strength and $\sigma_\mathcal{S}^z$ is the qubit Pauli matrix.
 Following Refs.~\cite{Quan2006,PFernandez2009}, we assume that
 the qubit is coupled with the environment only in the $|1\ra$ state.
 Then, according to Eq.~(\ref{SEH}), the effective Hamiltonian of the environment has the following expressions \cite{PFernandez2009}
 \bey
   \hat H_\mathcal{E}^{0}&=&-\frac{4(1-\alpha)}{N}\hat{S}_x^2+\alpha\hat{S}_z,
                     \label{EffH0} \\
  \hat H_\mathcal{E}^{1}&=&-\frac{4(1-\alpha)}{N}\hat{S}_x^2+(\alpha+\lambda)\hat{S}_z,
                     \label{EffH1}
 \eey
\noindent when the  qubit is in state $|0\ra$ and $|1\ra$, and where irrelevant constant terms have been ommited. It can be seen that the qubit  coupling to the environment is such that the two states of 
the qubit induce two different strengths in the second term of the environment Hamiltonian. 

 \begin{figure}
  \includegraphics[width=\columnwidth]{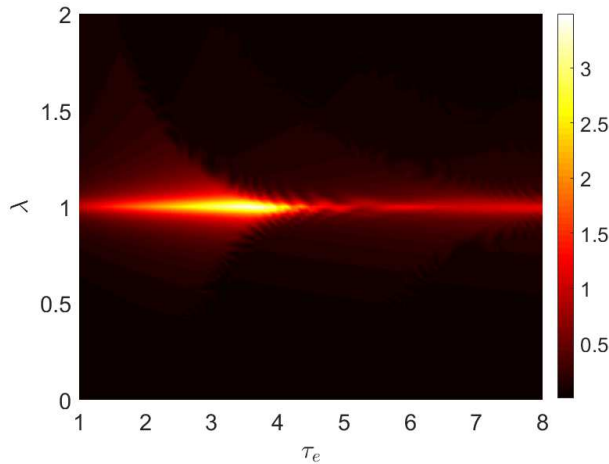}
  \caption{The quantum speed limit time, $\tau_\mathrm{QSL}$, as a function of the evolution
   time, $\tau_e$, and the coupling strength, $\lambda$,
   with a control parameter $\alpha=0.4$ and an  environment size $N=1000$.}
  \label{QSLDT}
 \end{figure}

 The variation of the coupling strength $\lambda$ can drive the
 environment through the critical point of the ground state quantum phase transition.  The
 interplay between the quantum criticality of the environment and the
 quantum speed limit time of the qubit has been analyzed in this case \cite{Wei2016,Hou2016} and it has been found that the ground state
 quantum phase transition of the environment substantially diminishes the speed of
 evolution of the qubit.  Our aim in the present work is, however, to assess the
 effect of the environment excited-state quantum phase transition on
 the quantum speed limit time of the qubit. 
 To this end, we need to know, for a
 control parameter $\alpha$ and critical energy $E_c$, the critical value of
 the coupling parameter --denoted as $\lambda_c$-- that will make the
 environment reach the critical energy $E_c$ of the excited-state quantum phase transition.

 In general, the critical energy of an excited-state quantum phase transition 
 is a function of the control parameter and
  the value of the critical coupling has to be obtained numerically for any initial state \cite{PFernandez2009,PFernandez2011}.  
 However, for the LMG bath (\ref{LMGH}), the critical energy is
 independent of the control parameter $\alpha$ as can be seen in Fig.\ \ref{Energys}.
 When the initial state of environment is the ground state,
 the critical value of the coupling $\lambda_c$ can be derived  using the intrinsic state formalism \cite{Relano2008}
 \be \label{LAMC}
    \lambda_c=2-\frac{5}{2}\alpha.
 \ee
We would like to emphasize  that the critical coupling $\lambda_c$, which
 induces the excited-state quantum phase transition in the environment Hamiltonian (\ref{LMGH}),
 differs from $\lambda_{c0}$, the critical coupling for the ground state quantum phase transition \cite{PFernandez2009,PFernandez2011}.

 \begin{figure*}
  \centering
  \includegraphics[width=0.8\textwidth]{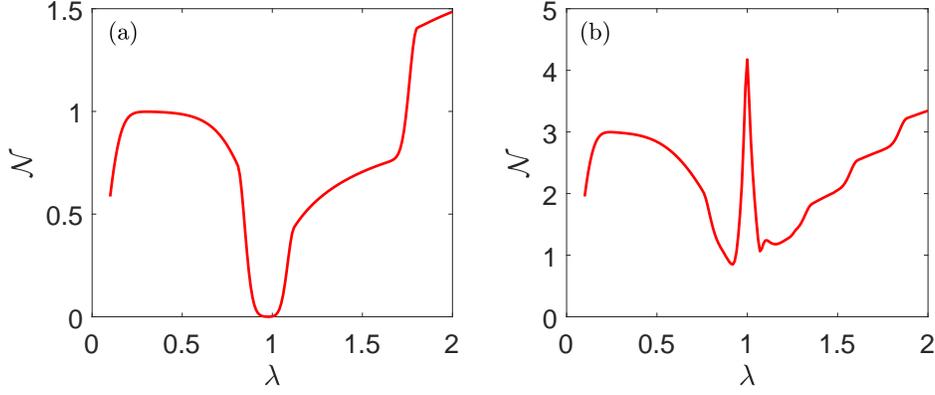}
  \caption{The non-Markovianity $\mathcal{N}$ as a function of coupling strength $\lambda$
   for $\alpha=0.4$ and environment size $N=1000$, with evolution time values $\tau_e=3.5$ (a)
   and $\tau_e=8$ (b).}
  \label{NvsLam}
 \end{figure*}

\section{Quantum speed limit time of the qubit} \label{Results}

 To understand the effects of the excited-state quantum phase transition on the quantum speed limit, 
we study the quantum speed limit time of the qubit after a quench 
with a coupling strength value such that
 it drives the environment  through the critical energy of the excited-state quantum phase transition. 
The initial state of the qubit is defined as
 $|\psi_0\ra=\cos(\theta/2)|0\ra+e^{-i\phi}\sin(\theta/2)|1\ra$, while the environment is assumed to be
 initially in the ground state of $\hat H_\mathcal{E}^0$: $|\Psi^0_{\mathcal{E},G}\ra$.
 Therefore, the wave function of the total system at $t=0$ is
 $|\Phi(0)\ra=[\cos(\theta/2)|0\ra+e^{-i\phi}\sin(\theta/2)|1\ra]\otimes|\Psi^0_{\mathcal{E},G}\ra$.
 The total system wave function evolves in time according to the Schr\"{o}dinger equation and
  the state of the total system at time $t$ is
 \be \label{Tstate}
   |\Phi(t)\ra=\cos(\theta/2)|0\ra\otimes|\Psi^0_{\mathcal{E}}(t)\ra
        +e^{-i\phi}\sin(\theta/2)|1\ra\otimes|\Psi^1_{\mathcal{E}}(t)\ra,
 \ee
 where the evolution of the environmental states $|\Psi^0_{\mathcal{E}}(t)\ra$ and $|\Psi^1_{\mathcal{E}}(t)\ra$
 satisfies the Schr\"{o}dinger equation
 $i\partial_t|\Psi^{l}_{\mathcal{E}}(t)\ra=\hat H_{\mathcal{E}}^{l}
     |\Psi_{\mathcal{E}}^{l}(t)\ra$,
 with $\hat H_\mathcal{E}^{l}$ for \(l = 0,1\) is given by Eqs.~(\ref{EffH0}) and (\ref{EffH1}), respectively.
 The evolution of the qubit, therefore, depends on the dynamics of the two environment branches evolving with
 effective Hamiltonians $\hat H_\mathcal{E}^0$ and $\hat H_\mathcal{E}^1$.

 From Eq.~(\ref{Tstate}), the reduced density matrix of the qubit in the $\{|0\ra,|1\ra\}$  basis at time $t$ is
 \begin{align}
  \rho_S(t)&=\mathrm{Tr}_{\mathcal{E}}|\Phi(t)\ra\la\Phi(t)| \notag \\
   &=
  \begin{bmatrix}
    \cos^2(\theta/2)  & \frac{1}{2}e^{i\phi}\sin(\theta)\mathcal{M}^\ast(t)  \\
    \frac{1}{2}e^{-i\phi}\sin(\theta)\mathcal{M}(t)  & \sin^2(\theta/2)
  \end{bmatrix},  \label{SystemD}
 \end{align}
 \noindent where
 $\mathcal{M}(t) =\la\Psi^0_{\mathcal{E},G}|e^{i\hat H_\mathcal{E}^{0}t}
 e^{-i\hat H_{\mathcal{E}}^{1}t}|\Psi^0_{\mathcal{E},G}\ra$ is the
 decoherence factor.  
 The modulus square of $\mathcal{M}(t)$ is known
 as the Loschmidt echo (LE), which has been widely studied in many
 fields (see, e.g., Ref.~\cite{Gorin2006} and references therein).  
 In the present work, we set the energy of the initial state
 $|\Psi^0_{\mathcal{E},G}\ra$ to zero and the decoherence factor
 $\mathcal{M}(t)$ in Eq.~(\ref{SystemD}) is reduced to
 \be \label{Dfactor}
 \mathcal{M}(t)=\la\Psi_{\mathcal{E},G}^0|e^{-i\hat H_{\mathcal{E}}^{1}t}|\Psi_{\mathcal{E},G}^0\ra,
 \ee
that can be identified with the survival probability of the initial
 quantum state \(|\Psi_{\mathcal{E},G}^0\ra\) evolving under the
 quenched Hamiltonian \(\hat H_{\mathcal{E}}^{1}\).

 In the rest of this section we investigate what happens to the 
 quantum speed limit time of the qubit when
 the environment undergoes an excited-state quantum phase transition.  
 We show how signatures of the excited-state quantum phase transition 
 manifest themselves in the quantum speed limit time of the qubit after
 quenching the coupling strength between qubit and environment in such
 a way that the environment passes through the critical point of the
 excited-state quantum phase transition.

 From Eq.~(\ref{SystemD}),  the
 quantum speed limit time (\ref{QSLT}) for the qubit evolution from $t=0$ to $t=\tau_e$ can be expressed as (see the appendix for the details)
 \be \label{QSLqubit}
    \tau_{\mathrm{QSL}}=\frac{\sin(\theta)\{1-\mathfrak{R}[\mathcal{M}(\tau_e)]\}}
                        {(1/\tau_e)\int_0^{\tau_e} dt|\partial_t\mathcal{M}(t)|},
 \ee
 where $\mathfrak{R}[\mathcal{M}(\tau_e)]$ denotes the real part of $\mathcal{M}(t)$ at $t=\tau_e$.
 In the following, without loss of generality, we set $\theta=\pi/2$.

 \begin{figure*}
  \includegraphics[width=\textwidth]{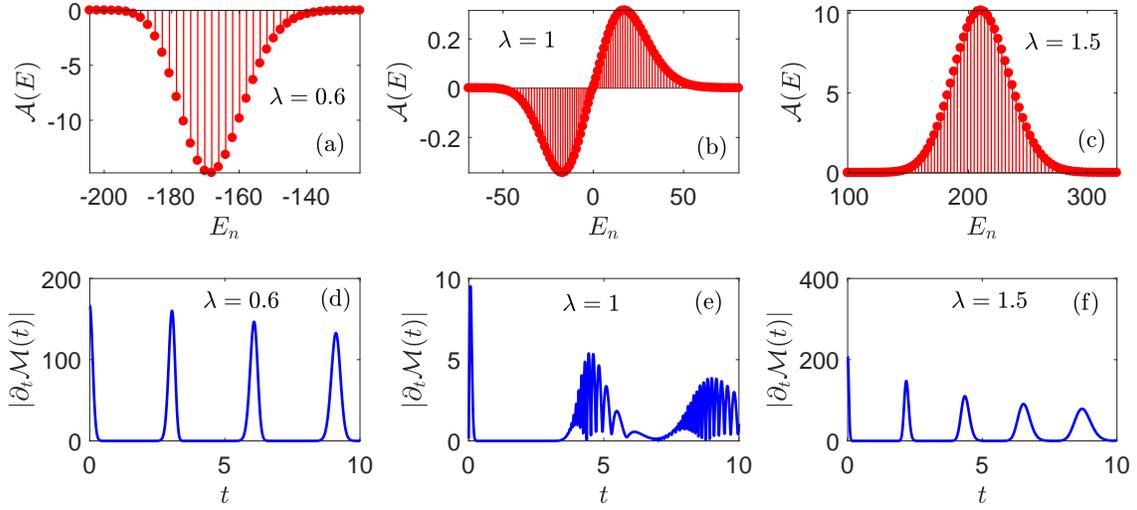}
  \caption{$\mathcal{A}(E)$ and $|\partial_t\mathcal{M}(t)|$
   for different coupling strengths for an environment with control parameter $\alpha=0.4$ and size $N=1000$.
   (a)-(c): $\mathcal{A}(E)$ as a function of $E_n$ [see Eq.~(\ref{SEgP})] with
   $\lambda$ below (a), at (b), and above (c) the critical coupling strength, respectively. Panels
   (d)-(f) represent the evolution of $|\partial_t\mathcal{M}(t)|$ calculated from Eq.~(\ref{dtM}) for the same coupling parameter values.}
  \label{EdM}
 \end{figure*}

 We depict in Fig.~\ref{QSLa}(a) the quantum speed limit time of the qubit as a
 function of the coupling strength $\lambda$. For the sake of
 simplicity, we set the evolution time $\tau_e=1$.  This figure has
 several remarkable features. On the one hand, when the value of the
 coupling strength is far away from the critical value $\lambda_c$,
 the quantum speed limit time is small. Moreover, increasing the environment size $N$,
 the evolution of the qubit can be accelerated.  On the other hand, it
 can be clearly noticed in the figure how the quantum speed limit time displays a
 sharp peak at the critical coupling value $\lambda_c$, which implies
 that driving the environment through the excite-state quantum phase transition 
 results in a slowdown of the quantum evolution of the qubit.  
The cusplike shape of the quantum speed limit time in the vicinity 
 of the critical coupling becomes sharper as the environment size $N$ increases.  
 Finally, the value of the quantum speed limit
 time at $\lambda_c$, i.e., the critical quantum speed limit time
 $\tau_\mathrm{QSL}(\lambda_c)$, is very close to the actual evolution
 time and is independent of the size of environment $N$.

 In Fig.~\ref{QSLa}(b) we show how the critical quantum speed limit time
 $\tau_{\mathrm{QSL}}(\lambda_c)$ changes with the environment size
 $N$, again for $\alpha=0.4$ and $\tau_e=1$.  It is obvious that
 $\tau_{\mathrm{QSL}}(\lambda_c)$ increases and tends to the actual
 evolution time for increasing $N$. By using the least square method, we find that the relation between $\tau_{\mathrm{QSL}}(\lambda_c)$ and $N$ is approximately $1-\tau_{\mathrm{QSL}}(\lambda_c)=N^{-\mu}$ [cf. the inset of Fig.~\ref{QSLa}(b)] with $\mu\approx1$. Therefore,
 $\tau_\mathrm{QSL}(\lambda_c)\to\tau_e$ in the thermodynamic limit
 $N\to\infty$.

 The features observed in the two panels of Fig.~\ref{QSLa} indicate that
 the quantum speed limit time can be used 
 as a proxy for excited-state quantum phase transitions in the environment.
 To further verify this statement, we compare the numerically
 estimated critical coupling strength, obtained as the location of the
 maximum in $\tau_\mathrm{QSL}$, with the exact value of $\lambda_c$
 in Eq.~(\ref{LAMC}).  The results are displayed in Fig.~\ref{NALmc},
 with a very good agreement between numerical and analytical results, evidencing
 that the qubit quantum speed limit time may be considered as 
 a reliable probe of excited-state quantum phase transitions
 in the LMG environment.

 The quantum speed limit time in Eq.~(\ref{QSLqubit}) also depends on the evolution time $\tau_e$.
 Figure~\ref{QSLDT} is a heatmap where we display the variation of the quantum speed limit time as a
 function of both $\tau_e$ and $\lambda$, with  $\alpha=0.4$ and $\lambda_c=1$.
 This figure has two remarkable features.
 On the first hand, for any $\tau_e$ value, the maximum $\tau_\mathrm{QSL}$ value  is always
 attained for the critical value of the coupling strength, $\lambda_c$.
 This feature confirms our previous suggestion of  the quantum speed limit time  
 as a probe for environment excited-state quantum phase transitions.
 On the second hand, the maximum value of $\tau_\mathrm{QSL}$ exhibits a nonmonotonic
 behavior for increasing evolution times. This is different from the ground state quantum phase transition case, 
 where the extreme  $\tau_\mathrm{QSL}$ value always 
 increases with the evolution time $\tau_e$ \cite{Wei2016}.

 \section{Mechanism for quantum slowdown of the qubit}  \label{PMc}

 Both theoretical \cite{Deffner2013b,Xu2014} and experimental \cite{Cimmarusti2015}
 studies have verified that
 the Markovian nature of the environment can slow down the quantum evolution of an open quantum system.
 Therefore, we need to investigate if there is any relationship between the increase of the 
 quantum speed limit   time due to the LMG environment excited-state quantum phase transition 
 and its Markovian nature, in order to
 clarify  the origin of the quantum slowdown mechanism 
 at the critical point of the excited-state quantum phase transition.

 For an open quantum system, the degree of Markovian behavior of the environment can be
 quantified by the measure of non-Markovianity, defined as \cite{Breuer2009}
 \be \label{NMD}
    \mathcal{N}=\underset{\rho_{1,2}(0)}{\mathrm{max}}\int_{\eta>0}\eta[t,\rho_{1,2}(0)]dt, 
    \ee
    
\noindent where the maximization is performed over all initial state pairs
 $\rho_{1,2}(0)$ and the integral is evaluated for all time intervals in which $\eta$ is positive. The integrand $\eta[t,\rho_{1,2}(0)]=d\{D[\rho_{1,2}(t)]\}/dt$ 
 is the time derivative of the trace distance 
$D[\rho_{1,2}(t)]=\mathrm{tr}|\rho_1(t)-\rho_2(t)|/2$ where
$\rho_{1,2}(t)$ is the time evolution of the initial $\rho_{1,2}(0)$.  
Here, $|\mathcal{O}|=\sqrt{\mathcal{O}^\dag \mathcal{O}}$ is
the trace norm for the $\mathcal{O}$ operator.

The distance $D$ defined above measures the distinguishability between
$\rho_{1}(t)$ and $\rho_{2}(t)$ and satisfies $0\leq D\leq1$.  It has
been demonstrated \cite{Breuer2009} that for Markovian dynamics all
states evolve towards a unique stationary state, which means
$\eta\leq 0$ and thus $\mathcal{N}=0$.  In Non-Markovian dynamics, due
to the information backflow from the environment to the system, the
distance $D$ can increase with time and $\eta$ is positive, which
leads to $\mathcal{N}>0$.

It has been found that, for the open system under consideration, the
optimal initial state pairs are provided by the equatorial, antipodal
states on the Bloch sphere \cite{Wissman2012, Haikka2012}.  Then, as
we show in the Appendix, $\mathcal{N}$ in Eq.~(\ref{NMD}) can be
rewritten as \cite{Wei2016}
\be
\label{CNM}
\mathcal{N}=\frac{1}{2}\left[\int_0^{\tau_e}|\partial_t|\mathcal{M}(t)||dt
  +|\mathcal{M}(\tau_e)|-1\right].
\ee

 In Fig.~\ref{NvsLam} we depict
 the measure of non-Markovianity $\mathcal{N}$ as a function of
 $\lambda$ with $\alpha=0.4$ and $N=1000$ for two
 different values of the evolution time. Comparing Fig.~\ref{NvsLam} with the behavior
 of quantum speed limit time depicted in Fig.~\ref{QSLDT}, one can clearly see that
 the qubit evolution slowdown at the critical point of the excited-state quantum phase transition is
 induced by the Markovian nature of the environment for short
 evolution times [Fig.~\ref{NvsLam}(a)], whereas for longer evolution times, both
 $\mathcal{N}$ and $\tau_\mathrm{QSL}$ exhibit a sharp peak at the
 critical point of the excited-state quantum phase transition as can be seen in Fig.~\ref{NvsLam}(b). 
 In the latter case the quantum evolution
 slowdown phenomenon cannot be
 explained exclusively from the Markovian nature of the environment
 and we need to reexamine the mechanism of the evolution slowdown
 phenomenon at the critical point of excited-state quantum phase transition.

 It is evident from Eq.~(\ref{QSLqubit}) that the quantum speed limit time depends on the 
 qubit decoherence rate through $|\partial_t\mathcal{M}(t)|$.
 Therefore, the evolution slowdown of the qubit at the critical point 
 of the excited-state quantum phase transition can be explained from
 the singular behavior of the decoherence rate.
 To verify this conjecture, in the following of this section we study the dynamics of $|\partial_t\mathcal{M}(t)|$.

 To this end, we take into consideration that the decoherence factor
 in Eq.~(\ref{Dfactor}) can be written as
 $\mathcal{M}(t)=\sum_ke^{-iE_kt}|c_k|^2=\int dE e^{-iEt}\omega(E)$,
 where $c_k=\la k|\Psi_{\mathcal{E},G}^0\ra$ denotes the expansion
 coefficient with $|k\ra$ of the $k$-th eigenstate of
 $H_\mathcal{E}^1$ and $E_k$ is the corresponding eigenenergy, while
 $\omega(E)=\sum_k|c_k|^2\delta(E-E_k)$, known as the strength
 function or the local density of states \cite{Santos2016}, is the
 energy distribution of the initial state $|\Psi_{\mathcal{E},G}^0\ra$
 in the eigenstates of $H_\mathcal{E}^1$.  We should emphasize that
 the strength function $\omega(E)$ has a complex shape at the critical
 point of the excited-state quantum phase transition, which leads to a
 characteristic behavior of the survival probability
 \cite{PFernandez2011,Santos2015,Santos2016,PBernal2017}.  Finally,
 the expression of the decoherence rate can be written as
 \be
 \label{dtM}
 |\partial_t\mathcal{M}(t)|=\left|\int dE
   e^{-iEt}\mathcal{A}(E)\right|,
 \ee
 where
 \be
 \label{SEgP}
 \mathcal{A}(E)=\sum_k|c_k|^2E_k\delta(E-E_k).
 \ee
 Evidently,
 $|\partial_t\mathcal{M}(t)|$ is the modulus of the Fourier transform
 of $\mathcal{A}(E)$ and its time dependence can be easily obtained
 from $\mathcal{A}(E)$.

 In Fig.~\ref{EdM}, we display $\mathcal{A}(E)$ and
 $|\partial_t\mathcal{M}(t)|$ for different coupling strengths with
 $\alpha=0.4$ and $N=1000$. Figs.~\ref{EdM}(a)-(c) show
 $\mathcal{A}(E)$ versus energy eigenvalues for coupling strength
 values below, at, and above the critical value, respectively.  For
 both non-critical coupling strengths $\mathcal{A}(E)$ is unimodal
 [see Figs.~\ref{EdM}(a) and \ref{EdM}(c)] and the peak location at
 $\la E\ra=\sum_k|c_k|^2 E_k$ depends on the value of $\lambda$.
 Moreover, the width of $\mathcal{A}(E)$ is approximately given by
 $\mathrm{Var}(E)=\sum_k|c_k|^2E_k^2-\la E\ra^2$.  However, the
 behavior of $\mathcal{A}(E)$ at the critical coupling value is more
 complex [see Fig.~\ref{EdM}(b)] with a double peak structure and
 negative (positive) $\mathcal{A}(E)$ values for negative (positive)
 energies and, therefore $\mathcal{A}(E_c)\approx0$ at the ESQPT
 critical energy $E_c=0$.

 The time evolution of $|\partial_t\mathcal{M}(t)|$ for the three
 coupling strengths discussed above is depicted in
 Figs.~\ref{EdM}(d)-(f).  In the first place, it can be easily noticed
 that the time dependence of $|\partial_t\mathcal{M}(t)|$, with
 regular damped oscillations, is qualitatively similar for
 non-critical values of $\lambda$ [see Figs.~\ref{EdM}(d) and
 \ref{EdM}(f)]. These oscillations, dependent on $\la E\ra$ and on the
 fine structure of $\mathcal{A}(E)$, have different frequencies
 and decay times. Furthermore, the characteristic time of the decaying
 envelope can be connected to the width of $\mathcal{A}(E)$ via a
 Heisenberg-like relation and the result is given by
 $\tau_c\propto1/\mathrm{Var}(E)$.  In the second place, there is a
 very different time dependence for $|\partial_t\mathcal{M}(t)|$ at
 the critical coupling [see Fig.~\ref{EdM}(e)], characterized by
 irregular oscillations. The regular damped recurrences displayed in
 Figs.~\ref{EdM}(d) and \ref{EdM}(f) are replaced by an initial sharp
 increase followed by groups of random oscillations.  Notice the
 different $y$-axis scaling in Figs.~\ref{EdM}(e-f), with
 $|\partial_t\mathcal{M}(t)|$ attaining minimum values for
 $\lambda=\lambda_c$.  The behavior of $|\partial_t\mathcal{M}(t)|$ at
 the critical point of the excited-state quantum phase transition can
 be traced back to the bimodal form of $\mathcal{A}(E)$ shown in
 Fig.~\ref{EdM}(b).

 In summary, we have shown that the existence of an 
 excited-state quantum phase transition in the LMG environment spectrum has a strong influence
 on the dynamics of the qubit, as can be seen from the time dependence of  $|\partial_t\mathcal{M}(t)|$. The critical value of the coupling strength leads the environment to 
 the excited-state quantum phase transition critical energy, where
 $|\partial_t\mathcal{M}(t)|$ displays a random oscillatory pattern of small amplitude.

\section{Conclusions}  \label{Final}

 In conclusion, we have analyzed the effects of an excited-state quantum phase transition on the quantum speed limit time of an open quantum system by studying
 a central qubit coupled to a spin environment modeled by a Lipkin-Meshkov-Glick bath.
 The signatures of the excited-state quantum phase transition in the environment are either  
 a divergence of the second order derivative of an individual excite state energy 
 with respect to the control parameter, or a singularity in the local density of states 
 at the critical energy ($E_c=0$) for a constant control  parameter.
 By quenching the coupling strength between the qubit and the environment,
 we have probed the impact that the excited-state quantum phase transition 
 in the environment has upon the quantum speed limit time of the qubit.

 We have found that the environment underlying excited-state quantum phase transition produces
 conspicuous effects in the quantum speed limit time of the qubit. 
 Namely, the quantum speed limit time of the qubit displays a sharp peak at the critical point of the
 excited-state quantum phase transition, making the excited-state quantum phase transition 
 responsible for a noticiable slow down in the qubit evolution.  
 Moreover, at the critical point of the excited-state quantum phase transition, the
 quantum speed limit time approaches the actual evolution time value for increasing
 environment size values.  
 In spite of the fact that long evolution times are
 associated with small values of the quantum speed limit time of the qubit, the
 maximum of the quantum speed limit time is always located at the critical energy of
 the excited-state quantum phase transition, making the quantum speed limit 
 time as a viable proxy for assessing the
 existence of an excited-state quantum phase transition in the LMG environment.

 With a calculation of a non-Markovianity measure for the open system
 under consideration, we have demonstrated that the qubit evolution
 slowdown at the critical point of excited-state quantum phase
 transition cannot be always explained from the Markovian nature of
 the environment.  In fact, the particular behavior of the quantum
 speed limit time at the critical point of the excited-state quantum
 phase transition stems from the singular behavior of
 $|\partial_t\mathcal{M}(t)|$ [cf.~Eq.~\ref{dtM}].  As happens with
 the LMG model, the different quantum many-body systems where
 excited-state quantum phase transitions have been identified exhibit
 a divergence in the state density at the critical point. This leads
 to a high localization for states at the critical point
 \cite{Santos2015,Santos2016, PBernal2017, Wang2017}, which makes the
 sum in Eq.~(\ref{SEgP}) giving as a result a small
 $\mathcal{A}(E)$. This means that $|\partial_t\mathcal{M}(t)|$ in
 Eq.~(\ref{dtM}) will oscillate in time with a small amplitude. As a
 consequence, $\tau_{\mathrm{QSL}}$ in Eq.~(\ref{QSLqubit}) has a
 sharp peak once the environment is located at its critical point.
 This indicates that  our analysis is independent of the LMG model details, and we conclude that the present results are valid in
 cases with an environment whose state density exhibits a divergence at
 the critical point of the excited-state quantum phase transition.
Furthermore, our work provides additional evidence that supports the
results of previous works that investigate the relation between the
quantum speed limit time and the Markovian character of the
environment (see, e.g., Ref.~\cite{Deffner2017} and references
therein).

The results reported in this work advance an original point of view on the influence of excited-state quantum phase transitions on quantum system dynamics. Furthermore, considering the recent experimental progresses on the detection of excited-state quantum phase transition signatures \cite{Dietz2013,Larese2011,Larese2013,Zhao2014},
 the quantum speed limit can be considered of an experimental proxy for excited-state quantum phase transitions in the laboratory.

\acknowledgments

Q.~W. gratefully acknowledges support from the National Science Foundation of China
under grant No. 11805165.
F.~P.~B. contribution to this work was partially funded by MINECO grant FIS2014-53448-C2-2-P, by the Consejería de 
Conocimiento, Investigación y Universidad, Junta de Andalucía and 
European Regional Development Fund (ERDF), ref. SOMM17/6105/UGR, and by the CEAFMC at the Universidad de Huelva.
The authors would like to thank Lea Santos for  useful discussions and suggestions.

\begin{appendix}

\section{Derivations of Eqs.~(\ref{QSLqubit}) and (\ref{CNM})}

In this appendix we derive Eqs.~ (\ref{QSLqubit}) and (\ref{CNM}).

\subsection{Derivation of equation (\ref{QSLqubit})}
For an initial pure state, the Bures angle (\ref{BAD}) can be written as \cite{Deffner2013b,Jozsa1994}
\be
   \mathcal{L}(\rho_0,\rho_t)=\arccos\sqrt{\mathrm{tr}(\rho_0\rho_t)}.
\ee
 Therefore, we have $\sin^2\mathcal{L}(\rho_0,\rho_t)=1-\mathrm{tr}[\rho_t\rho_0]$.
From the evolved density matrix of the qubit Eq.~(\ref{SystemD}), we find
 \begin{align}
   \mathrm{tr}[\rho_t\rho_0]=
    1-\frac{1}{2}\sin^2\left(\theta\right)\{1-\mathfrak{R}[\mathcal{M}(t)]\},
 \end{align}
where $\mathfrak{R}[\mathcal{M}(t)]$ denotes the real part of  $\mathcal{M}(t)$.
So, we obtain
 \begin{align}  \label{SnL}
  \sin^2\mathcal{L}(\rho_0,\rho_t)
     =\frac{1}{2}\sin^2\left(\theta\right)\{1-\mathfrak{R}[\mathcal{M}(t)]\}. 
 \end{align}
 Substituting Eq.~(\ref{SystemD}) into Eq.~(\ref{Dyopen}), we find that the Liouvillian superoperator $L_t(\rho_t)$ has following expression
 \begin{align}
   L_t(\rho_t)&=\dot{\rho}(t)  \notag \\
   &=
   \begin{pmatrix}
   0 &  (e^{i\phi}/2)\sin\left(\theta\right)\dot{\mathcal{M}}^\ast(t) \\
   (e^{-i\phi}/2)\sin\left(\theta\right)\dot{\mathcal{M}}(t) & 0
   \end{pmatrix},
 \end{align}
where the dot denotes the time derivative.
 Obviously, $L_t^\dag(\rho_t)=L_t(\rho_t)$, therefore, the singular values of $L_t$ are given by the
 absolute value of the eigenvalues of $L_t(\rho_t)$
 \be 
   l_1=l_2=\frac{1}{2}\sin\left(\theta\right)|\partial_t \mathcal{M}(t)|.
 \ee
Then the Schatten $p$ norm of $L_t(\rho_t)$ reads
\be
   ||L_t(\rho_t)||_p=(l_1^p+l_2^p)^{1/p}=2^{1/p}\frac{\sin(\theta)|\partial_t \mathcal{M}(t)|}{2}.
\ee
Hence, we have $||L_t(\rho_t)||_\infty<||L_t(\rho_t)||_2<||L_t(\rho_t)||_1$. As a result,
\be \label{Sgv}
   \mathrm{max}\left\{\frac{1}{\Gamma^1_{\tau_e}}, \frac{1}{\Gamma^2_{\tau_e}},
   \frac{1}{\Gamma^\infty_{\tau_e}}\right\}=\frac{1}{\Gamma^\infty_{\tau_e}}
  =\dfrac{2/\sin(\theta)}{(1/\tau_e)\int_0^{\tau_e}|\partial_t\mathcal{M}(t)|dt}.
\ee
 Substituting Eqs.~(\ref{SnL}) and (\ref{Sgv}) into the expression of the quantum speed limit time Eq.~(\ref{QSLT}), we thus obtain
 \be
   \tau_{QSL}=\dfrac{\sin\left(\theta\right)\{1-\mathfrak{R}[\mathcal{M}(\tau_e)]\}}
              {(1/\tau_e)\int_0^{\tau_e}|\partial_t\mathcal{M}(t)|dt}.
 \ee

\subsection{Derivation of Eq.~(\ref{CNM})}

For the open system we studied in this work, it has been verified that the optimal initial pair 
$\rho_{1,2}(0)$ in Eq.~(\ref{NMD}) are given by equatorial ($\theta=\pi/2$), antipodal states \cite{Wissman2012, Haikka2012}.
Therefore, we have
\begin{align}
  \rho_1(0)=
  \begin{pmatrix}
   1/2  &   e^{i\phi}/2  \\
   e^{-i\phi}/2  & 1/2
  \end{pmatrix},  
  \rho_2(0)=
  \begin{pmatrix}
   1/2  &   -e^{i\phi}/2  \\
   -e^{-i\phi}/2  & 1/2
  \end{pmatrix}.
\end{align}
Then, the evolution of $\rho_{1,2}(0)$ is given by
\begin{align}
  &\rho_1(t)=
  \begin{pmatrix}
   1/2  &  (e^{i\phi}/2)\mathcal{M}^\ast(t)  \\
   (e^{-i\phi}/2)\mathcal{M}(t)  & 1/2
  \end{pmatrix}, \\ 
  &\rho_2(t)=
  \begin{pmatrix}
   1/2  &   -(e^{i\phi}/2)\mathcal{M}^\ast(t)  \\
   -(e^{-i\phi}/2)\mathcal{M}(t)  & 1/2
  \end{pmatrix}.
\end{align}
We thus have
\be
  \rho_1(t)-\rho_2(t)=
  \begin{pmatrix}
   0  &   e^{i\phi}\mathcal{M}^\ast(t)  \\
   e^{-i\phi}\mathcal{M}(t)  & 0
  \end{pmatrix}.
\ee
So, we obtain
\be
 |\rho_1(t)-\rho_2(t)|=
  \begin{pmatrix}
   |\mathcal{M}(t)|  &   0  \\
   0  & |\mathcal{M}(t)|
  \end{pmatrix}.
\ee
The trace distance $D[\rho_1(t),\rho_2(t)]$ is, therefore, given by
\be
  D[\rho_1(t),\rho_2(t)]=\frac{1}{2}\mathrm{tr}|\rho_1(t)-\rho_2(t)|
        =|\mathcal{M}(t)|.
\ee

Now, we have $\eta=\partial_t|\mathcal{M}(t)|$ and 
the measure of non-Markovianity in Eq.~(\ref{NMD}) reads
\be
  \mathcal{N}=\int_{\eta>0}\partial_t|\mathcal{M}(t)|dt.
\ee
Noting that
\begin{align}
  &\int\partial_t|\mathcal{M}(t)|dt=
    \int_{\eta>0}\partial_t|\mathcal{M}(t)|dt+\int_{\eta<0}\partial_t|\mathcal{M}(t)|dt, \notag \\
  &\int|\partial_t|\mathcal{M}(t)||=
    \int_{\eta>0}\partial_t|\mathcal{M}(t)|dt-\int_{\eta<0}\partial_t|\mathcal{M}(t)|dt,
\end{align}
and employing $\int_0^{\tau_e}\partial_t|\mathcal{M}(t)|dt=|\mathcal{M}(\tau_e)|-1$, we 
finally obtain 
 \be
   \mathcal{N}=\frac{1}{2}\left[\int_0^{\tau_e}|\partial_t|\mathcal{M}(t)||dt+|\mathcal{M}(\tau_e)|-1\right].
 \ee

\end{appendix}

\bibliographystyle{apsrev4-1}
\bibliography{QSLtex_3}

\end{document}